\documentclass{article}
\usepackage[utf8]{inputenc}
\usepackage{graphicx}
\usepackage{rotating}
\usepackage{subfig}
\usepackage{amsmath}
\usepackage{cite}
\title{Control of Permanent Magnet Motors with Actuation Bounds using Convex Optimization}
\author{Shravan Mohan\\\small \textit{Santa Fe Research, Chennai}}
\date{August 2019}

\begin{document}

\maketitle

\begin{abstract}
    \noindent This paper presents a nonlinear control algorithm for speed control of a open-end winding  permanent magnet motor. The idea relies on a feedback linearization technique which also ensures adherence to current and voltage bounds. These bounds arise from practical limitations of the power source. The feedback linearization law is computed using a convex optimization routine to minimize response time as well. The aid of convex optimization leads to computational efficiency. Moreover, the mathematical tractability of the approach also aids analysis of the system performance under model uncertainty and feedback measurement noise. Simulations and computations corroborate the proposed idea.
\end{abstract}

\section{An Introduction}
A permanent magnet motor is a device which converts electrical energy to mechanical energy, by generating spatially and temporally varying magnetic fields \cite{fitzgerald2003electric}. The permanent magnet, simply stating, consists of the stator, or the outer casing, and the rotor. The stator comprises of multiple wound coils with open ends (an open-end winding topology will be considered through out this paper) and the rotor comprises of permanent magnets firmly attached to its shaft. The varying magnetic field is generated by passing current through the stator coils. The misalignment in the net magnetic field generated by the stator coil and the net magnetic field of the permanent magnets on the rotor gives rise to torque on the shaft. This is the principle on which the motor rotates.\\\\
The natural question is then that of determining appropriate currents through the stator coils to effect rotation of the motor at a desired speed. The currents in turn are generated by applying voltages across the open ends of the coils. It is conceivable that the voltages required must be within practical limits of the voltage source. Hence, the currents must also be limited due to the power limitation of the voltage source and also for safe operation. As for designing the control algorithm, the difficulty arises from the non-linearity of the dynamical system model of the motor \cite{sastry2013nonlinear}. In brief, the non-linearity stems from the fact that the torque developed on the rotor is proportional to a non-linear function of the rotor angle with respect to a stationary frame. Note that the rotor angle itself is a state-space variable. Therefore, the bounds on actuation only complicate the problem by limiting the feasible control set.\\\\
A popular paradigm in the area of non-linear control is that of feedback linearization. The basic idea is to device a feedback function of the state-space variables such that the resulting system is rendered linear. The questions of existence of the such a function and its construction are well studied in literature  \cite{sastry2013nonlinear}. However, as will be seen in this paper, the approach will not depend explicitly on these constructs, although a implicit dependence may be found. Instead, the existence and the construction will be ascertained by a solution to a set of linear equations,  with convex Linear Matrix Inequality (LMI) constraints ensuring actuation bounds  \cite{boyd2004convex}.\\\\
The field of speed control has been an active area of research and the following references, although seminal, do not constitute a comprehensive coverage of the literature. For example, the authors of  \cite{bodson1993high} present a feedback linearized nonlinear controller based on  feedback linearization. Their idea uses the well known DQ transformation for converting the canonical model of a stepper motor to one which has polynomial non-linearity. The resulting ODE is then linearized by cancelling the non-linearities using full-state feedback. The feedback law is also tailored to adhere to a voltage constraint, while also maximizing the torque produced at the desired speed. Moreover, a speed observer is also devised for cases where only currents and position are known. In \cite{moehle2015optimal}, the authors present a method based on convex optimization to develop a current waveform which minimizes the torque ripple and the resistive energy lost in the coils. The proposed method also takes into account the general nonlinearity in the torque function \& back-emfs, and current/voltage bounds, but the route is not through a feedback linearization. Instead, the functional optimization problem is solved by discretisation, which leads to a more tractable convex constrained quadratic program. It might also be possible to devise a controller using the technique presented in this paper to include costs of resistive power dissipation. In \cite{vsmidl2017direct}, the authors present an optimal control algorithm based on local linearization of the nonlinear dynamical system model. Since the power dissipated is a quadratic functional, they develop an optimal linear state feedback law by solving the Ricatti equation. However, since the solution is not guaranteed to lie within actuation bounds, another optimization problem to obtain the closest optimal solution to the unconstrained optimal waveform, which also adheres to bounds and the dynamic constraints. In the present paper, however, a linearizing feedback controller is devised with a general nonlinear  model and the currents \& voltages being bounded. Moreover, current measurements are not required for feedback linearization and the controlled is constructed for minimum response time. To the best of the authors' knowledge, the proposed method has not been presented earlier. The main contributions of the paper are the following:
\begin{itemize}
    \item The model takes into account the distortions (other than the expected sinusoidal variation) in the torque generated on the rotor (as a function of the rotor angle) by the stator currents. These distortions may be result of defects in construction.
    \item The construction of the control methodology is simple to understand and  computationally efficient for real-time implementations.
    \item The method can be easily extended to the following cases: (i) heterogeneous torque functions for different stator coils, (ii) faulty and hence non-conducting stator coils, and (iii) permanent magnet motors with more than three phases.
    \item Since the system is feedback linearized, certain performance guarantees can be given under model uncertainties and feedback measurement noise. 
\end{itemize}
This paper is organized as follows. The second section will outline the dynamical system model for the standard open-end winding three-phase permanent magnet motor. The third section will discuss the construction of the feedback linearization functional. The fourth section will provide for a way to incorporate the actuation bounds on voltages and currents through LMIs. The fifth section will discuss the performance of the control strategy under model uncertainty and feedback measurement noise. The last section will present simulation results corroborating the proposed idea, followed by concluding remarks.
\section{The Dynamical System}
\begin{figure}[t]
    \centering
    \includegraphics[scale=0.25]{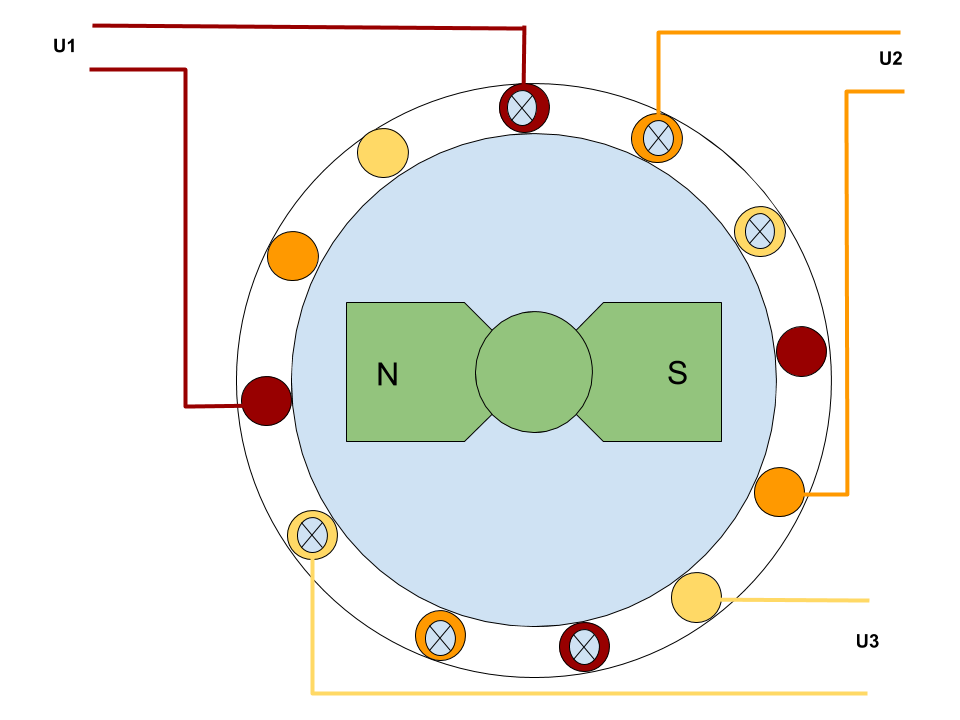}
    \caption{A simplified model of the stator and the rotor of a Permanent Magnet motor. This is a cross-sectional view of the motor. The angle that the center rotor makes with the horizontal axis is the state space element $\boldsymbol{\theta}$. Differently colored coils are spatially separated by $2\pi/3$ radians. The cross marks indicate that the current goes inwards into the plane.}
    \label{fig:motor_schematic}
\end{figure}

The conventional dynamical system model for a three-phase permanent magnet motor is given by:
\begin{equation} 
    \begin{array}{l}
         \displaystyle \boldsymbol{\dot{\boldsymbol{\theta}}} = \boldsymbol{\boldsymbol{\omega}}  \\
         \displaystyle J\boldsymbol{\dot{\boldsymbol{\omega}}} = \left(\overline{m}_r\times \overline{m}_s\right)\overline{e}_z - \boldsymbol{T_{in}} \\
        \displaystyle  L\boldsymbol{\dot{i}_1} + \boldsymbol{i_1}R = \boldsymbol{\boldsymbol{u_1}} - \boldsymbol{\boldsymbol{\omega}} \boldsymbol{\mathrm{b}}(\boldsymbol{\boldsymbol{\theta}}) \\
        \displaystyle  L\boldsymbol{\dot{i}_2} + \boldsymbol{i_2}R = \boldsymbol{\boldsymbol{u_2}} - \boldsymbol{\boldsymbol{\omega}} \boldsymbol{\mathrm{b}}\left(\boldsymbol{\boldsymbol{\theta}}+\frac{2\pi}{3}\right) \\
        \displaystyle  L\boldsymbol{\dot{i}_3} + \boldsymbol{i_3}R = \boldsymbol{\boldsymbol{u_3}} - \boldsymbol{\boldsymbol{\omega}} \boldsymbol{\mathrm{b}}\left(\boldsymbol{\boldsymbol{\theta}}+\frac{4\pi}{3}\right),
    \end{array}
    \label{eqn:dynsys}
\end{equation}
where the parameters of the aforementioned model are given by: 
\begin{equation} 
    \begin{array}{l}
         \displaystyle \overline{m}_s = \overline{m}_1 + \overline{m}_2 + \overline{m}_3\\
        \displaystyle  \overline{m}_1 = K\boldsymbol{i_1}\overline{e}_0 ,~
         \displaystyle \overline{m}_2 = K\boldsymbol{i_2}\overline{e}_{\frac{2\pi}{3}} ,~
        \displaystyle  \overline{m}_3 = K\boldsymbol{i_3}\overline{e}_{\frac{4\pi}{3}} \\
        \displaystyle \boldsymbol{\mathrm{b}}(\boldsymbol{\boldsymbol{\theta}}) = \sum_{k=1}^{N}c_k\cos(k\boldsymbol{\boldsymbol{\theta}}) +s_k\sin(k\boldsymbol{\boldsymbol{\theta}}) \\
        \displaystyle  \overline{m}_r = P\overline{e}_{\boldsymbol{\boldsymbol{\theta}}}\\
        \displaystyle  -U \leq \boldsymbol{\boldsymbol{u_1}},~\boldsymbol{\boldsymbol{u_2}},~\boldsymbol{\boldsymbol{u_3}} \leq U \\
         \displaystyle -A \leq \boldsymbol{i_1},~\boldsymbol{i_2},~\boldsymbol{i_3} \leq A.
    \end{array}
    \label{eqn:params}
\end{equation}
Here in \eqref{eqn:dynsys}, $\boldsymbol{\boldsymbol{\theta}}$ is the angular position of the rotor with respect to the fixed horizontal axis and $\boldsymbol{\boldsymbol{\omega}}$, the angular speed. Refer to Figure \ref{fig:motor_schematic} for a schematic of a typical permanent magnet motor. The first equation in \eqref{eqn:dynsys} arises from the definition of angular speed. The second equation in \eqref{eqn:dynsys} governs the rate of change of the angular velocity. In that, the first term is the torque between the net magnetic fields created by the stator and the rotor, \textit{i.e.}, $\overline{m}_r$ and $\overline{m}_s$, respectively. The final term $\boldsymbol{T_{in}}$ is an external torque on the rotor shaft, which is assumed to be known apriori. In a 3-phase motor, the stator is made up for 3 identical coils, which generate magnetic fields separated angularly by $2\pi/3$ radians, given by $\overline{e}_0, \overline{e}_{\frac{2\pi}{3}}, \overline{e}_{\frac{4\pi}{3}}$ in \eqref{eqn:params}. Since magnetic field vectors  interfere additively, the net magnetic field generated by the stator is the vector sum of those generated by the three coils. This fact gives rise to the first equation in \eqref{eqn:params}. The magnitude of the magnetic field generated by each stator coil is assumed to be linearly proportional to the current passing through it, the  proportionality constant being $K$. These are mathematically described by the second, third and the fourth equations in \eqref{eqn:params}. Now, since each of the stator coils has an inductance and resistance, the dynamics of the currents is dictated by an appropriate linear differential equation, given by the third, fourth and the fifth differential equations in \eqref{eqn:dynsys}. There is also the back-emf created in the stator coils as the rotor rotates. This is dependent on the position of the rotor, as well as its angular velocity. It is assumed that the back-emf is  directly proportional to the angular speed, and its dependence on $\boldsymbol{\boldsymbol{\theta}}$ is assumed to be known apriori. The fifth equation in \eqref{eqn:params} governs this back-emf. 
Suppose that $\boldsymbol{\boldsymbol{u_1}}, \boldsymbol{\boldsymbol{u_2}}, \boldsymbol{\boldsymbol{u_3}}$ are the control voltages applied to the stator coils. These also appear in the differential equations governing the currents in \eqref{eqn:dynsys}. These are the inputs to the system. The rotor magnetic field is in the direction of $\boldsymbol{\boldsymbol{\theta}}$ and assumed to have a constant magnitude $P$. This fact is reflected in the sixth equation of \eqref{eqn:params}. Finally, the control voltages have actuation bounds due to practical limitations of the voltage source. Similarly, the current in the coils cannot exceed certain bounds for safe operation. These bounds typically depends on several factors such as (i) the material of the coil, (ii) the insulation material used and (iii) the maximum shear stress  that the rotor shaft can sustain. These constitute the last inequalities in \eqref{eqn:params}. \\\\
The aforementioned model , however, is not the most general due to the following main reason. The torque developed on the rotor by the stator currents at a particular angular displacement is proportional to the sine of the angle. But, the complicated magnetic field distribution inside the stator core might lead to distortions in the torque generated. In other words, the dependence of torque on the rotor angle might have distortions other than pure sinusoid. Call this function $\boldsymbol{\mathrm{f}}(.)$. A similar argument then holds for the back-emf produced while the rotor rotates. Denote this function $\boldsymbol{\mathrm{g}}(.)$. Since the functions are still periodic, one can write them as a Fourier series. The usefulness of this representation will become clear in the next section. The dynamical system can be written in a more conventional way as the following. Note that here $\boldsymbol{\boldsymbol{x_1}}$ is the rotor angle, $\boldsymbol{\boldsymbol{x_2}}$ the angular velocity, and $\boldsymbol{\boldsymbol{x_3}}, \boldsymbol{x_4}$ and $\boldsymbol{x_5}$ the three phase currents. The torque produced by any coil when the current through it is $\boldsymbol{\boldsymbol{x_3}}$ and the rotor angle is $\boldsymbol{\boldsymbol{x_1}}$ is given by $\boldsymbol{\boldsymbol{x_3}}\boldsymbol{\mathrm{f}}(\boldsymbol{\boldsymbol{x_1}})$. Similar expressions hold for the other two  coils as well, just that the arguments are incremented by $2\pi/3$ and $4\pi/3$ radians, respectively.
\noindent \begin{equation}
    \begin{array}{l}
         \displaystyle \dot{\boldsymbol{\boldsymbol{x_1}}} ~=~ \boldsymbol{\boldsymbol{x_2}}  \\
         \displaystyle \dot{\boldsymbol{\boldsymbol{x_2}}} ~=~ \boldsymbol{\boldsymbol{x_3}}\boldsymbol{\mathrm{f}}(\boldsymbol{\boldsymbol{x_1}}) + \boldsymbol{x_4}\boldsymbol{\mathrm{f}}\left(\boldsymbol{\boldsymbol{x_1}}+\frac{2\pi}{3}\right) + \boldsymbol{x_5}\boldsymbol{\mathrm{f}}\left(\boldsymbol{\boldsymbol{x_1}}+\frac{4\pi}{3}\right) - \boldsymbol{T_{in}}\\
        \displaystyle  \dot{\boldsymbol{\boldsymbol{x_3}}} ~=~ \boldsymbol{u_1} - \tau \boldsymbol{\boldsymbol{x_3}} - \boldsymbol{\boldsymbol{x_2}} \boldsymbol{\mathrm{g}}(\boldsymbol{\boldsymbol{x_1}}) \\
        \displaystyle  \dot{\boldsymbol{x_4}} ~=~ \boldsymbol{u_2} - \tau \boldsymbol{x_4} - \boldsymbol{\boldsymbol{x_2}} \boldsymbol{\mathrm{g}}\left(\boldsymbol{\boldsymbol{x_1}}+\frac{2\pi}{3}\right) \\
        \displaystyle  \dot{\boldsymbol{x_5}} ~=~ \boldsymbol{u_3} - \tau \boldsymbol{x_5}- \boldsymbol{\boldsymbol{x_2}} \boldsymbol{\mathrm{g}}\left(\boldsymbol{\boldsymbol{x_1}}+\frac{4\pi}{3}\right). 
    \end{array}
\end{equation}
\\
\noindent \underline{\textbf{Remark on Identification of System Parameters}}: \textit{It is also necessary to discuss the identification of the functions $\boldsymbol{\mathrm{f}}(.)$ and $\boldsymbol{\mathrm{g}}(.)$, although it is assumed in this paper that these are known apriori. The function $\boldsymbol{\mathrm{f}}(.)$ can be determined under no load conditions. Suppose that one of the coils is energized with a DC voltage source and the initial rotor position is at $\pi/2$ radians relative to the net magnetic field produced by that coil. While the  rotor aligns to the stable position with respect to the stator magnetic field, the currents, the rotor angle and the angular acceleration can be measured. The same procedure can be done with initial rotor position at $3\pi/2$ radians, and energizing the other coils. Now, with this data, the function $\boldsymbol{\mathrm{f}}(.)$ can be inferred from the angular acceleration. On the other hand, the function $\boldsymbol{\mathrm{g}}(.)$ can be estimated by rotating the rotor with an external torque and measuring the emf generated across the coils.}
\section{The Feedback Linearization}
The feedback linearization, as mentioned earlier, is a method of devising a feedback which renders the system linear. This way controlling the system into performing a desired behaviour becomes much easier. To begin with, it is assumed that the rotor angle and the rotor angular velocity are both measured and are thus available as feedback signals. Also, the following simplifications is useful. Note that the back-emf in the current equations of \eqref{eqn:dynsys} can be subsumed into the control signals themselves. In other words, $\boldsymbol{u_1}+\boldsymbol{x_2}g(\boldsymbol{\boldsymbol{x_1}})$ can be termed as $\boldsymbol{u_1}$, with some abuse of notation but without any loss of generality. Similar simplifications can be made for $\boldsymbol{u_2}$ and $\boldsymbol{u_3}$. Reverting is also easy, \textit{i.e.}, once the currents are known as functions of the state variables, the actual inputs $\boldsymbol{u_1}$, $\boldsymbol{u_2}$ and $\boldsymbol{u_3}$ can be determined easily. Therefore, the system reduces to a second order non-linear dynamical system with three inputs, all being the currents in the stator coils. Mathematically, the reduced system is given by:
    \begin{equation}
        \begin{array}{l}
         \dot{\boldsymbol{\boldsymbol{x_1}}} = \boldsymbol{x_2}\\
         \dot{\boldsymbol{\boldsymbol{x_2}}} = \boldsymbol{\boldsymbol{u_1}} \boldsymbol{\mathrm{f}}(\boldsymbol{\boldsymbol{x_1}}) + \boldsymbol{u_2} \boldsymbol{\mathrm{f}}(\boldsymbol{\boldsymbol{x_1}}+2\pi/3) + \boldsymbol{u_3} \boldsymbol{\mathrm{f}}(\boldsymbol{\boldsymbol{x_1}}+4\pi/3) - \boldsymbol{T_{in}}. 
        \end{array}
        \label{eqn:redsys}
    \end{equation}
The basic idea here will be easily understood with the help of the following example. Assume that $\boldsymbol{\mathrm{f}}(\boldsymbol{x}) = \sin(\boldsymbol{x})$. In that case, suppose one chooses the currents as $\boldsymbol{\boldsymbol{u_1}} = \textbf{I}\sin(\boldsymbol{x_1})$, $\boldsymbol{u_2} = \textbf{I}\sin\left(\boldsymbol{x_1} + \frac{2\pi}{3}\right)$ and $\boldsymbol{u_3} = \textbf{I}\sin\left(\boldsymbol{x_1} + \frac{4\pi}{3}\right)$.  Then the equation for torque reads, post substitution for the currents as:
    \begin{equation}
        \begin{array}{l}
             J\dot{\boldsymbol{\omega}} = 1.5 P K \textbf{I} - \boldsymbol{T_{in}}.
        \end{array}
    \end{equation}
The simplification of the right hand side of the above equation happens due to the following trigonometric identity:
$$
\sin^2(\boldsymbol{\theta}) + \sin^2(\boldsymbol{\theta}+2\pi/3)+ \sin^2(\boldsymbol{\theta}+4\pi/3) = 3/2. 
$$
This implies that for ensuring that the speed settles to $\boldsymbol{\boldsymbol{\omega}_r}$, one can set $$\displaystyle \textbf{I} = \frac{\boldsymbol{T_{in}}}{1.5PK}(\boldsymbol{\boldsymbol{\omega}_r} - \boldsymbol{\omega} + 1).$$ 
In this case, the system settles to the desired speed just as a first order stable linear system would.  To obtain the control voltage $\boldsymbol{u_1}$, the expression for $I$ can be substituted  into the ordinary differential equation for the current in \eqref{eqn:dynsys}, and get:
    \begin{equation}\footnotesize
        \begin{array}{l}
           \displaystyle
           \boldsymbol{u_1} = \boldsymbol{\omega} \boldsymbol{\mathrm{b}}(\boldsymbol{x_1}) +  
          \frac{\boldsymbol{T_{in}}}{1.5PK}\left((\boldsymbol{\boldsymbol{\omega}_r} - \boldsymbol{\omega} + 1)(R\cos(\boldsymbol{x_1}) + \boldsymbol{\omega}L\sin(\boldsymbol{x_1})) + \frac{L\boldsymbol{T_{in}}}{J}(\boldsymbol{\boldsymbol{\omega}_r}-\boldsymbol{\omega})\sin(\boldsymbol{\boldsymbol{x_1}})\right).
        \end{array}
    \end{equation}
The other two control voltages $\boldsymbol{u_2}$ and $\boldsymbol{u_3}$ can be obtained similarly. Crucially, one still has to take care of the bounds on the current and the control voltages, which will be the topic of discussion for the next section. \\\\
The take away from the previous example was the fact that trigonometric functions aided the construction of the linearizing feedback for trigonometric force fields. Thus, the motivation now is to use a trigonometric approximation  of the non-linear torque-rotor angle dependence. This approximation is obtained using a truncated Fourier series of the function $\boldsymbol{\mathrm{f}(.)}$. Of course, such an approximation dilutes the model, but that will be dealt with by providing performance guarantees under model uncertainties in a later section. Therefore, suppose
\begin{equation}
    \begin{array}{l}
         \displaystyle \boldsymbol{\mathrm{f}}(\boldsymbol{\theta}) = \sum_{k=1}^{M}p_k\cos(k\boldsymbol{\theta}) +q_k\sin(k\boldsymbol{\theta})
    \end{array}
\end{equation}
As was done in the previous example, suppose the currents are given by the expressions:
    \begin{equation}
        \begin{array}{l}
             \displaystyle \boldsymbol{u_1} = \sum_{k=0}^M p^{(1)}_k \cos(k\boldsymbol{\theta}) + q^{(1)}_k \sin(k\boldsymbol{\theta}),  \\
             \displaystyle \boldsymbol{u_2} = \sum_{k=0}^M p^{(2)}_k \cos(k\boldsymbol{\theta}) + q^{(2)}_k \sin(k\boldsymbol{\theta}),  \\
             \displaystyle \boldsymbol{u_3} = \sum_{k=0}^M p^{(3)}_k \cos(k\boldsymbol{\theta}) + q^{(3)}_k \sin(k\boldsymbol{\theta}).  
        \end{array}
    \end{equation}
Substituting these expressions for current, one can obtain the total torque generated on the rotor by the stator coils as:
\begin{equation}
    \begin{array}{cc}
        \displaystyle \textbf{I} + \sum_{k=1}^{2M} \mathcal{C}_k\left(p^{(1)}_{\{1,\cdots,M\}}, q^{(1)}_{\{1,\cdots,M\}},p^{(2)}_{\{1,\cdots,M\}}, q^{(2)}_{\{1,\cdots,M\}},p^{(3)}_{\{1,\cdots,M\}}, q^{(3)}_{\{1,\cdots,M\}}\right) \cos(k\boldsymbol{\theta}) \\ + \displaystyle \mathcal{S}_k\left(p^{(1)}_{\{1,\cdots,M\}}, q^{(1)}_{\{1,\cdots,M\}},p^{(2)}_{\{1,\cdots,M\}}, q^{(2)}_{\{1,\cdots,M\}},p^{(3)}_{\{1,\cdots,M\}}, q^{(3)}_{\{1,\cdots,M\}}\right) \sin(k\boldsymbol{\theta}) , 
        \end{array}
    \end{equation}
    where all coefficients given by $\mathcal{C}_k$s and $\mathcal{S}_k$s are linear affine functionals of their respective arguments. Also note that the number of harmonics in the expression are twice the number of harmonics used for approximating $\boldsymbol{\mathrm{f}}(.)$. The crucial idea now is to choose these such that all these functionals are identically zero. This leaves only the term $I$ which can be set to 1, without loss of generality. In fact, if all the coefficients defining the control inputs defining the currents are scaled by
    $$
    \left(\boldsymbol{\boldsymbol{x^*_2}} - \boldsymbol{\boldsymbol{x_2}} + \boldsymbol{\boldsymbol{T_{in}}}\right),
    $$
    the resulting first order linear ODE is stable and therefore would settle to the desired angular speed $x^*_2$. The requirements on the value of $\mathcal{C}_k$s and $\mathcal{S}_k$s translate to a set of linear equations on the coefficients of the control input. As will be seen later, consistency of this system of linear equations ensures the existence of a control strategy within actuation bounds. Also note that the consistency of such a system can be checked easily.
    \\\\
    \underline{\textbf{Remark on Existence of Solutions}}: \textit{It is also necessary shed light on the question of conditions on $p_k$'s and $q_k$'s which lead to a consistency system of linear equations. In particular, this leads to the following problem in linear algebra: Under what conditions does a non-zero vector $[c^*_M~\cdots ~c_1^*~0~c_1~\cdots ~c_M]$ allow  vector $[0~\cdots~0~1~0~\cdots~0]^\top$ to be in the union of the column spaces of the following matrices: 
    $$
    A = \begin{pmatrix}
        c_M & \\
        c_{M-1} & c_M\\
        \vdots \\
        c^*_M & \cdots & c_1^* & 0 &c_1 & \cdots & c_M\\
        & & & & & & \vdots\\
        & & & & & c_M^* & c_{M-1}^*\\
        & & & & & &  c_M^*
    \end{pmatrix},
    $$
    $$
    B = \begin{pmatrix}
        c_M z^{M} & \\
        c_{M-1}z^{M-1} & c_Mz^{M}\\
        \vdots \\
        c^*_Mz^{-M} & \cdots & c_1^*z^{-1} & 0 &c_1z^1 & \cdots & c_Mz^{M}\\
        & & & & & & \vdots\\
        & & & & &  c_M^*z^{-M} & c_{M-1}^*z^{-(M-1)}\\
        & & & & & & c_M^*z^{-M}
    \end{pmatrix},
    $$
    and 
    $$
    C = \begin{pmatrix}
        c_M z^{2M} & \\
        c_{M-1}z^{2(M-1)} & c_Mz^{2M}\\
        \vdots \\
        c^*_Mz^{-2M} & \cdots & c_1^*z^{-2} & 0 &c_1z^{2} & \cdots & c_Mz^{2M}\\
        & & & & & & \vdots\\
        & & & & &  c_M^*z^{-2M} & c_{M-1}^*z^{-2(M-1)}\\
        & & & & & & c_M^*z^{-2M}
    \end{pmatrix}.
    $$
    Note that the number of variables is $6M+3$, while the total number of equations to be satisfied is $4M+1$. This fact makes the existence of a solution at least plausible. If it is shown that there exists one solution, there would be infinitely many solutions. With some algebra, one can easily show that there are no solutions to this system of equations if all non-triplen coefficients are zero. However, this case implies that the magnetic field repeats thrice in every full rotation. This means that there is no need for three separate coils. Therefore, it will be henceforth assumed that at least one of the non-triplen coefficients is non-zero. It has been observed through simulations that the union of column spaces of the aforementioned matrices spans the whole range space. This observation needs further investigation.    
    \\\\
    \noindent A couple of additional points need mention here. Firstly, the model as such does not take care of mutual inductance. But note that this translates to just another application of a linear transformation. Secondly, the magnetic fields produced by coils need not be symmetric (discounting the $2\pi/3$ radian shifts). The method extends naturally to that case as well. In fact, the angular shift in coils also need not be $2\pi/3$ for this method to be applicable.  Moreover, it is also easy to see that the method also extends to motors with more than 3 poles. This translates to having more force terms in the differential equation governing the angular acceleration, and more control inputs. Secondly, no measurement is devoid of noise. Therefore, the feedback must also be noisy. This aspect will be discussed after the construction of the control algorithm and certain performance guarantees will also be given.}
\section{The Actuation Bounds}
So far, the linearizing feedback has been characterized by a system of linear equations. Moreover, there are infinitely many solutions. The next step is to choose solutions which adhere to current and voltage constraints. While choosing a solution within the desired bounds, one must also ensure that the closed loop response is as fast as possible for optimal performance. This requirement, as we shall see, will lead to a convex optimization problem. \\\\
Firstly, consider the current constraints. Note that the currents are given by real expressions of the form:
\begin{equation}
    r_0 + \sum_{k=0}^{k=N} r_k e^{j k \boldsymbol{\theta}}+ r_k^* e^{-j k \boldsymbol{\theta}}.
\end{equation}
The above quantity is lesser than $I_{max}$ if and only if there exists a symmetric matrix $Q_{\max}$:
\begin{equation}
    \begin{array}{l}
         \begin{pmatrix}
            Q_{\max} + AQ_{\max}A^\top & AQ_{\max}B^\top \\
            BQ_{\max}A^\top & 0
         \end{pmatrix} + 
         \begin{pmatrix}
            0 & C^\top \\
            C & \frac{1}{2}\left(D+D^\top\right)
         \end{pmatrix} \succeq 0,
    \end{array}
\end{equation}
where, 
\begin{equation}
    \begin{array}{l}
         A = \begin{pmatrix}
            0 & I\\
            I & 0
         \end{pmatrix}, B = [0, \cdots, 1], C = -[r_1, \cdots, r_N]^\top ~\&~ D =  \displaystyle \frac{I_{\max}-r_0}{2}. 
    \end{array}
\end{equation}
Similarly, the current is more than $\displaystyle I_{\min}$ if and only if there exists a symmetric matrix $Q_{\min}$:
\begin{equation}
    \begin{array}{l}
         \begin{pmatrix}
            Q_{\min} + AQ_{\min}A^\top & AQ_{\min}B^\top \\
            BQ_{\min}A^\top & 0
         \end{pmatrix} + 
         \begin{pmatrix}
            0 & C^\top \\
            C & \frac{1}{2}\left(D+D^\top\right)
         \end{pmatrix} \succeq 0,
    \end{array}
\end{equation}
where, 
\begin{equation}
    \begin{array}{l}
         A = \begin{pmatrix}
            0 & I\\
            I & 0
         \end{pmatrix}, B = [0, \cdots, 1], C = [r_1, \cdots, r_N]^\top ~\&~ D =  \displaystyle \frac{r_0-I_{\min}}{2}. 
    \end{array}
\end{equation}
In the above definitions, $I$ is the identity matrix and the dimensions are chosen appropriately. Note that the above matrix inequalities depend only linearly (or affinely) on the unknowns. This implies, that the current constraints are essentially LMIs which are convex in the optimization variables.  \\\\
The voltage constraints can be dealt in the following way. First of all, the voltage bound $V_{\min} \leq V\leq V_{\max}$ implies that $I_{\min}/R \leq I \leq V_{\max}/R$, where $R$ is the coil resistance of the motor. Typically, a required voltage waveform is generated as an appropriate switched waveform using an inverter (this will be outline in the next remark). This switched waveform when fed to the motor coils gets low pass filtered naturally due to the high inductance of the coils. Hence the current generated in the coils as a result is the one desired, while the voltage bounds are trivially satisfied. It is worth reiterating that the voltage waveform can be easily computed using the dynamical system once the currents are known. In the simulations presented in this paper, it is assumed that the voltage source limits are much larger than the computed quantity. Note that computed voltage also has the back-emf adding up. However, if one wants to incorporate it into the optimization, a different mathematical formulation can be used. This would also be a part of a separate discussion in later remark.  \\\\
\underline{\textbf{Remark on Generation of Switching Control Voltages:}} \textit{Suppose the current signal to be generated in a coil is given by the continuous function $i(t)$. Also assume that the coil has a inductance of $L$ and a resistance of $R$. Moreover, the switched mode voltage supply can generate two levels given by $\{-V, +V\}$. It is immediately clear that $i(t)$ must satisfy: $-V/R \leq i(t) \leq V/R, ~\forall ~t$. Also suppose that $|\dot{i}(t)| \leq M$. Now, suppose a triangle-PWM is generated from the voltage source, by comparing $i(t)$ with a triangular wave whose time-period is lesser than $M/10$. This voltage waveform when filtered by the $RL$-impedance of the coil will generate a current $\Tilde{i}(t)$ which will be very close to the desired $i(t)$. Of course, the difference between $\tilde{i}(t)$ and $i(t)$ can be made smaller by higher frequency triangle wave, but then this would also be limited by the highest switching frequency of the devices in the switched mode voltage supply.}\\\\ 
\underline{\textbf{Remark on Generating Continuous Control Voltages:}} \textit{As mentioned earlier, in the case one wants to incorporate the back-emf calculations, a different mathematical formulation is needed. To that end, once the currents are known as functions of the rotor angle and angular speed, the control voltage can be obtained using their respective governing dynamical equations (recall that the voltages appear in the differential equations governing coil currents). With some simple algebraic substitutions and normalizing the angular speed to the interval $[0,1]$, one can note that the control voltage will be given by expressions of the form:
\begin{equation}
    \boldsymbol{\omega}^2 F_1(\boldsymbol{\theta}) + \boldsymbol{\omega} F_2(\boldsymbol{\theta}) + F_3(\boldsymbol{\theta}).
\end{equation}
The bounds on voltage naturally translates to 
\begin{equation}
    V_{\min} \leq \boldsymbol{\omega}^2 F_1(\boldsymbol{\theta}) + \boldsymbol{\omega} F_2(\boldsymbol{\theta}) + F_3(\boldsymbol{\theta}) \leq V_{\max}.
\end{equation} 
Consider $\boldsymbol{\omega}^2 F_1(\boldsymbol{\theta}) + \boldsymbol{\omega} F_2(\boldsymbol{\theta}) + \left(F_3(\boldsymbol{\theta})-V_{\min}\right)$. Since such an expression must be positive for each $0\leq \boldsymbol{\omega} \leq 1$, this is true if and only if:
\begin{equation}
    \begin{array}{l}
         F_3 - V_{\min} \geq 0, ~\forall \boldsymbol{\theta} \\
         F_2 + 2\left(F_3- V_{\min}\right) \geq 0, ~\forall \boldsymbol{\theta} \\
         F_1 + F_2 + F_3 - V_{\min} \geq 0, ~\forall \boldsymbol{\theta}. 
    \end{array}
\end{equation}
The above result stems from the fact that a polynomial is positive over $[0,1]$ if and only if the coefficients in its Bernstein polynomial basis representation are all non-negative. Then, the above three trigonometric constraints can be further translated to their respective convex LMIs, just as was done for the current constraints. A similar set of LMI constraints can also be derived for the case of upper bound on voltages. These LMI added to the main optimization problem would then have to be solved.}\\\\ 
With all this knowledge, one can then formulate an optimization problem so as to achieve the best possible motor performance. This essentially means that the control must be designed to achieved desired speed equilibrium as fast as possible. Mathematically, this boils down to maximizing the constant term generated while the cosines and sines terms are eliminated in the feedback linearization process. The optimization problem, which is convex, is shown in \eqref{eqn:mainalgo}.
\begin{figure}[t]
\centering
\fbox{\begin{minipage}{35em}
\begin{equation}
\begin{array}{l}
    \max_{p^{(1)},q^{(1)},p^{(2)},q^{(2)},p^{(3)},q^{(3)}} ~~~~t\\
    \mbox{subject to}\\
    \hspace*{1cm} \mathcal{C}_k\left(p^{(1)},q^{(1)},p^{(2)},q^{(2)},p^{(3)},q^{(3)}\right) = 0, ~\forall 1\leq k \leq 2N,\\
    \hspace*{1cm} \mathcal{S}_k\left(p^{(1)},q^{(1)},p^{(2)},q^{(2)},p^{(3)},q^{(3)}\right) = 0, ~\forall 1\leq k \leq 2N,\\
    \hspace*{1cm} \mathcal{C}_0\left(p^{(1)},q^{(1)},p^{(2)},q^{(2)},p^{(3)},q^{(3)}\right) + \mathcal{S}_0\left(p^{(1)},q^{(1)},p^{(2)},q^{(2)},p^{(3)},q^{(3)}\right) = t,\\\\
    \hspace*{2cm}\begin{pmatrix}
            Q_{\max} + AQ_{\max}A^\top & AQ_{\max}B^\top \\
            BQ_{\max}A^\top & 0
         \end{pmatrix} + 
         \begin{pmatrix}
            0 & C^\top \\
            C & \frac{1}{2}\left(D_1+D_1^\top\right)
    \end{pmatrix} \succeq 0,\\\\
    \hspace*{2cm}\begin{pmatrix}
            Q_{\min} + AQ_{\min}A^\top & AQ_{\min}B^\top \\
            BQ_{\min}A^\top & 0
         \end{pmatrix} + 
         \begin{pmatrix}
            0 & C^\top \\
            C & \frac{1}{2}\left(D_2+D_2^\top\right)
    \end{pmatrix} \succeq 0,\\\\
    \mbox{where},~~A = \begin{pmatrix}
            0 & I\\
            I & 0
         \end{pmatrix},~~ B = [0, \cdots, 1],~~ C = -[r_1, \cdots, r_N]^\top,\\\\ \mbox{and~~}D_1 =  \displaystyle \frac{I_{\max}-r_0}{2} ~\&~  D_2 = \displaystyle \frac{r_0-I_{\min}}{2}.
\end{array}
\label{eqn:mainalgo}
\end{equation}
\end{minipage}}
\caption{The main optimization problem for determining the control coefficients.}
\label{fig:mainalgo}
\end{figure}
\section{Model Uncertainties \& Noise}
The case for incorporating model uncertainties and feedback measurement noise is quite important. Firstly, the system identification process might introduce some errors in system parameters. Secondly, the truncation of the Fourier series of the torque function of coils also introduces model uncertainty. Finally, no measurement device is free of noise, which leads to erroneous feedback. These uncertainties will be assumed to bounded; this assumption is certainly fair for measurement noise, as practical measurement devices cannot have an error distribution with infinite support.\\\\
Mathematically, the effect of these perturbations is tractable as feedback linearization helps reduce the model to a first order stable linear differential equation in angular speed, and the perturbations are assumed to be bounded. Now, suppose that the linear differential equation governing angular speed post feedback linearization is given by:
\begin{equation}
    \dot{\omega} = K(\omega_r - \omega) + d(t),
\end{equation}
where $d(t)$ is the net perturbation. To elaborated on $d(t)$, consider the case where the feedback is given by $x_1 + \epsilon_1$ and $x_2 + \epsilon_2$. In that case, the two differential equation governing $x_2$ can be written as:
\begin{equation}\tiny
    \begin{array}{l}
        \displaystyle \dot{\boldsymbol{\boldsymbol{x_2}}} ~=~ \left(x_2^* - x^2 + T_{in} + \eta_{x_2} + \eta_{T_{in}}\right)\left(\left(\sum_{k=0}^M p^{(1)}_k \cos(k\boldsymbol{\theta}) + q^{(1)}_k \sin(k\boldsymbol{\theta}) + \eta^1_{\boldsymbol{\theta}}\right)\left(\sum_{k=1}^{M}p_k\cos(k\boldsymbol{\theta}) +q_k\sin(k\boldsymbol{\theta}) + \boldsymbol{\tilde{f}}(\boldsymbol{\theta})\right)\right. \\\displaystyle
        \hspace*{0cm}\displaystyle+\left(\sum_{k=0}^M p^{(2)}_k \cos(k\boldsymbol{\theta}) + q^{(2)}_k \sin(k\boldsymbol{\theta})+ \eta^2_{\boldsymbol{\theta}}\right)\left(\sum_{k=1}^{M}p_k\cos(k(\boldsymbol{\theta}+2\pi/3)) +q_k\sin(k(\boldsymbol{\theta}+2\pi/3)) + \boldsymbol{\tilde{f}}(\boldsymbol{\theta}+2\pi/3)\right)\\\displaystyle\left.
        \hspace*{0cm}\displaystyle+\left(\sum_{k=0}^M p^{(3)}_k \cos(k\boldsymbol{\theta}) + q^{(3)}_k \sin(k\boldsymbol{\theta})+\eta^3_{\boldsymbol{\theta}}\right)\left(\sum_{k=1}^{M}p_k\cos(k(\boldsymbol{\theta}+4\pi/3)) +q_k\sin(k(\boldsymbol{\theta}+4\pi/3)) + \boldsymbol{\tilde{f}}(\boldsymbol{\theta}+4\pi/3)\right)\right) - T_{in}
    \end{array}
\end{equation}
With some algebra, one can write this as:
\begin{equation}
    \begin{array}{l}
        \displaystyle
      \dot{x}_2(t)  = x_2^* - x^2 + d(t),
    \end{array}
\end{equation}
where,
\begin{equation}\tiny 
    \begin{array}{l} 
        \displaystyle
        d(t) = \left(x_2^* - x_2  + T_{in}+\eta_{x_2} + \eta_{T_{in}}\right)\left(\eta^1_\theta \tilde{f}(\boldsymbol{\theta}) + \eta^2_\theta \tilde{f}(\boldsymbol{\theta}+2\pi/3)+ \eta^3_\theta \tilde{f}(\boldsymbol{\theta}+4\pi/3) +  \tilde{f}(\boldsymbol{\theta})\left( \sum_{k=0}^N p_k^1\cos(k\boldsymbol{\theta}) + q_k^1\sin(k\boldsymbol{\theta}) \right)\right. \\\displaystyle+ \tilde{f}(\boldsymbol{\theta}+2\pi/3)\left( \sum_{k=0}^N p_k^2\cos(k\boldsymbol{\theta}) + q_k^2\sin(k\boldsymbol{\theta}) \right) \displaystyle + \tilde{f}(\boldsymbol{\theta}+4\pi/3)\left( \sum_{k=0}^N p_k^3\cos(k\boldsymbol{\theta}) + q_k^3\sin(k\boldsymbol{\theta}) \right)\\
        \displaystyle + \eta^1_{\boldsymbol{\theta}}\left( \sum_{k=0}^N p_k\cos(k\boldsymbol{\theta}) + q_k\sin(k\boldsymbol{\theta}) \right) + \eta^2_{\boldsymbol{\theta}}\left( \sum_{k=0}^N p_k\cos(k\boldsymbol{\theta}+2\pi/3) + q_k\sin(k\boldsymbol{\theta}+2\pi/3) \right) \\\displaystyle\left. + \eta^3_{\boldsymbol{\theta}}\left( \sum_{k=0}^N p_k\cos(k\boldsymbol{\theta}+4\pi/3) + q_k\sin(k\boldsymbol{\theta}+4\pi/3) \right)\right) + \eta_{x_2} + \eta_{T_{in}}.
    \end{array}
\end{equation}
Since each term in the above expression is bounded, the perturbation is also bounded. That is $|d(t)|\leq \eta$, for some $\eta$. The exact calculation of $\eta$ can be made from the aforementioned formula. Note that here it was also assumed that the speed ($x_2$) is always within the rating of the machine. It is also assumed that the perturbation is continuous with respect to time. Now, Gronwall's inequality states the following. Suppose $\leq -Kx(t) - b(t) \leq \dot{x}(t) = -Kx(t) + \eta(t) \leq -Kx(t) + b(t)$, and that $x_1(t)$, $x_2(t)$ are solutions to $\dot{x}(t) = -Kx(t) - b(t)$ and $\dot{x}(t) = -Kx(t) + b(t)$, respectively. Then $x_1(t)\leq x(t)\leq x_2(t)$, over any finite time interval $[t_0, t_1]$. In that case, after sufficiently long time (as $t\rightarrow \infty$), the desired angular speed would be off by at most $\eta/K$ rads/sec from the desired value of $\omega_r$.
\section{Simulations \& Computations}
\begin{sidewaysfigure}
  \centering
  \subfloat{\includegraphics[scale=0.45]{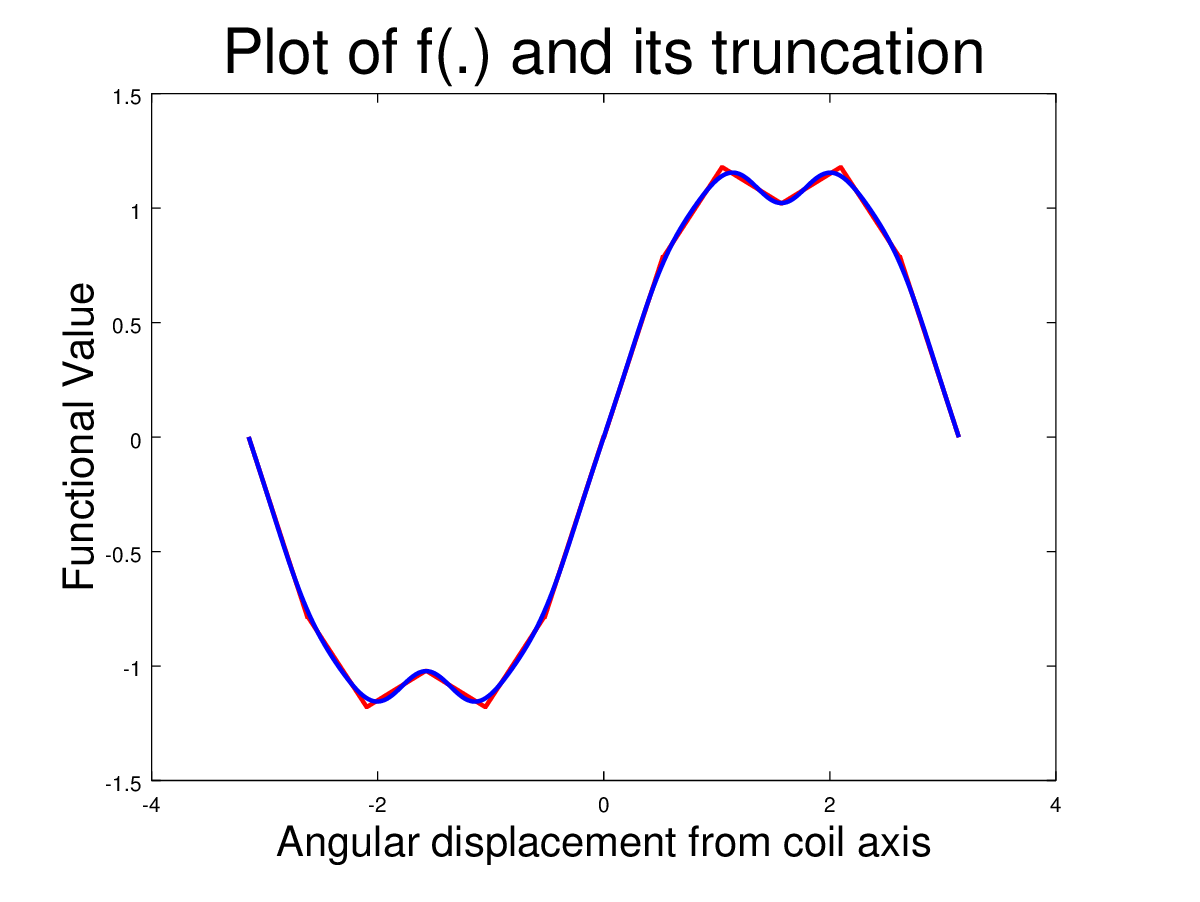}} \quad
  \subfloat{\includegraphics[scale=0.45]{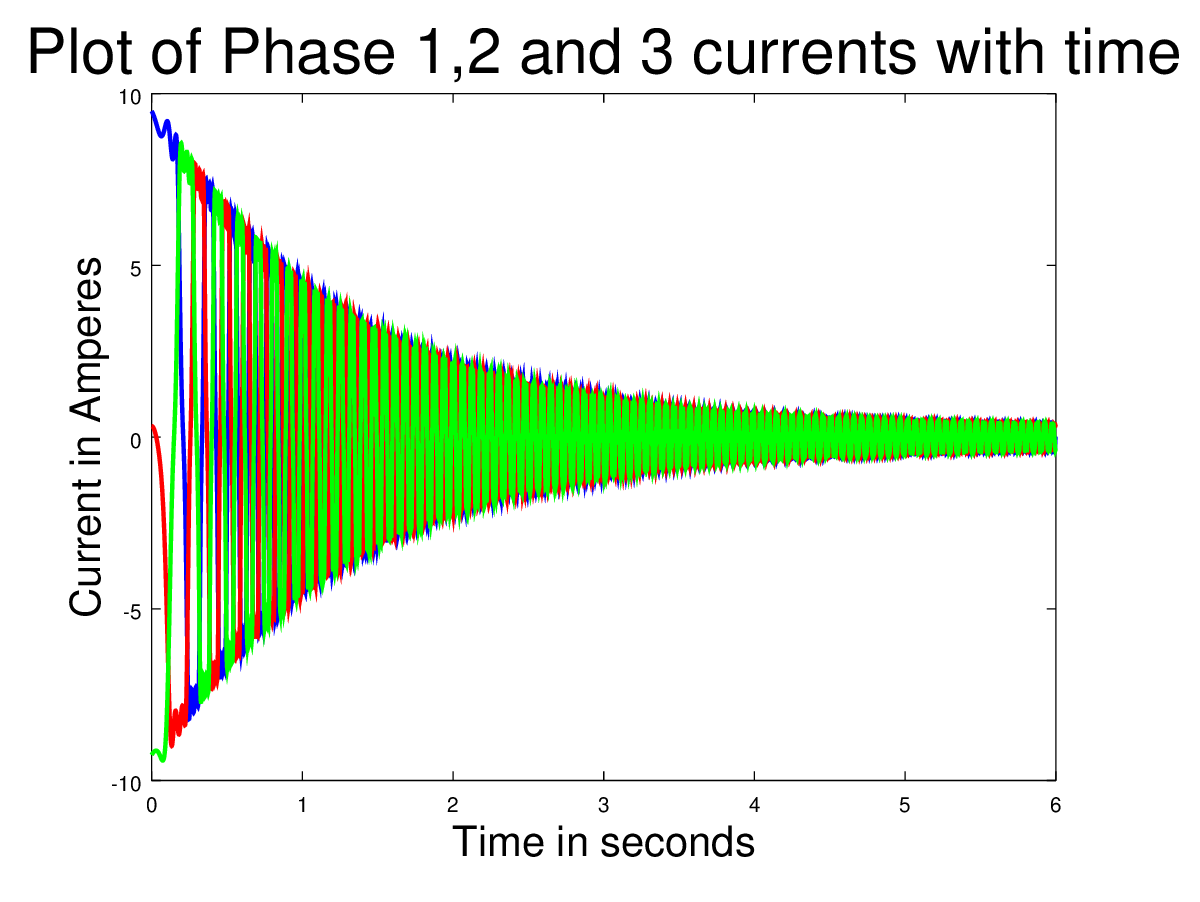}}\quad
  \subfloat{\includegraphics[scale=0.45]{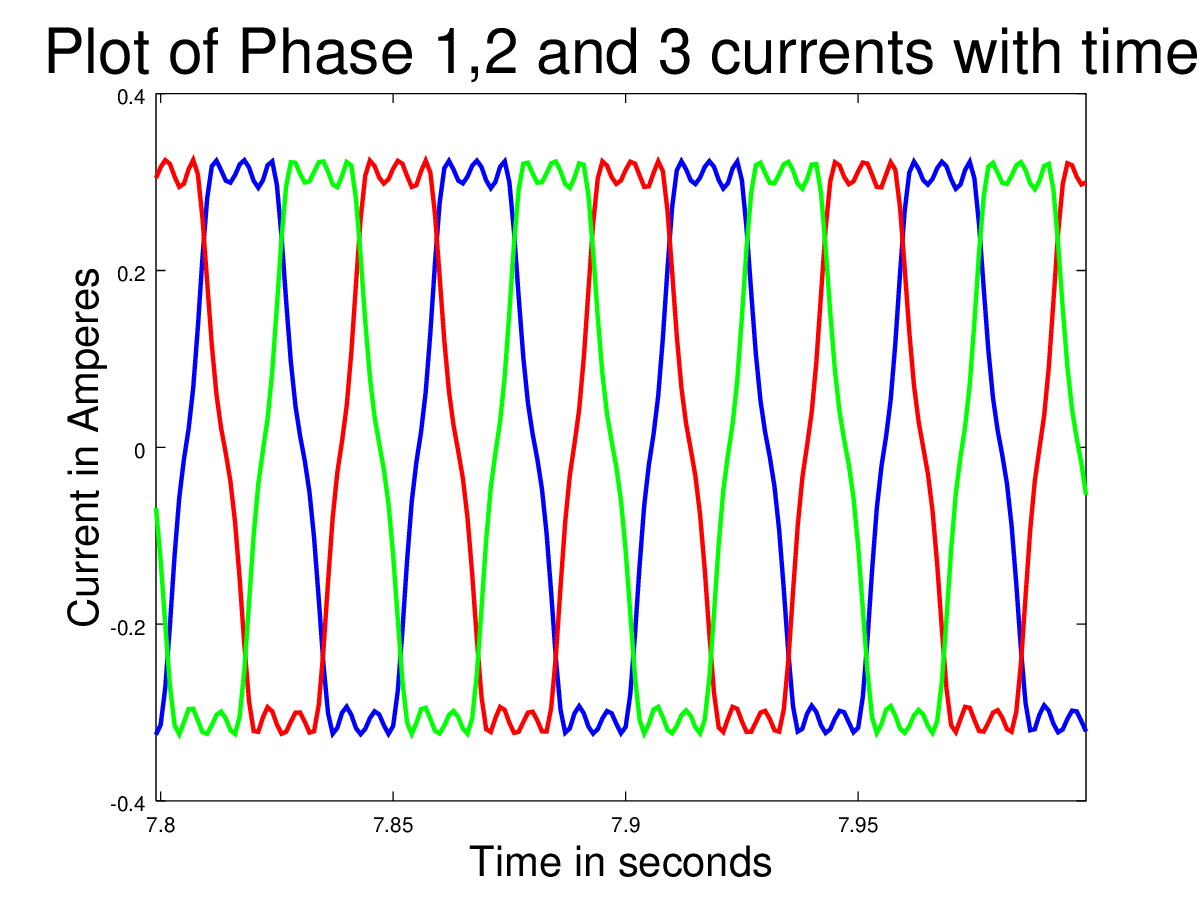}}\quad
  \subfloat{\includegraphics[scale=0.45]{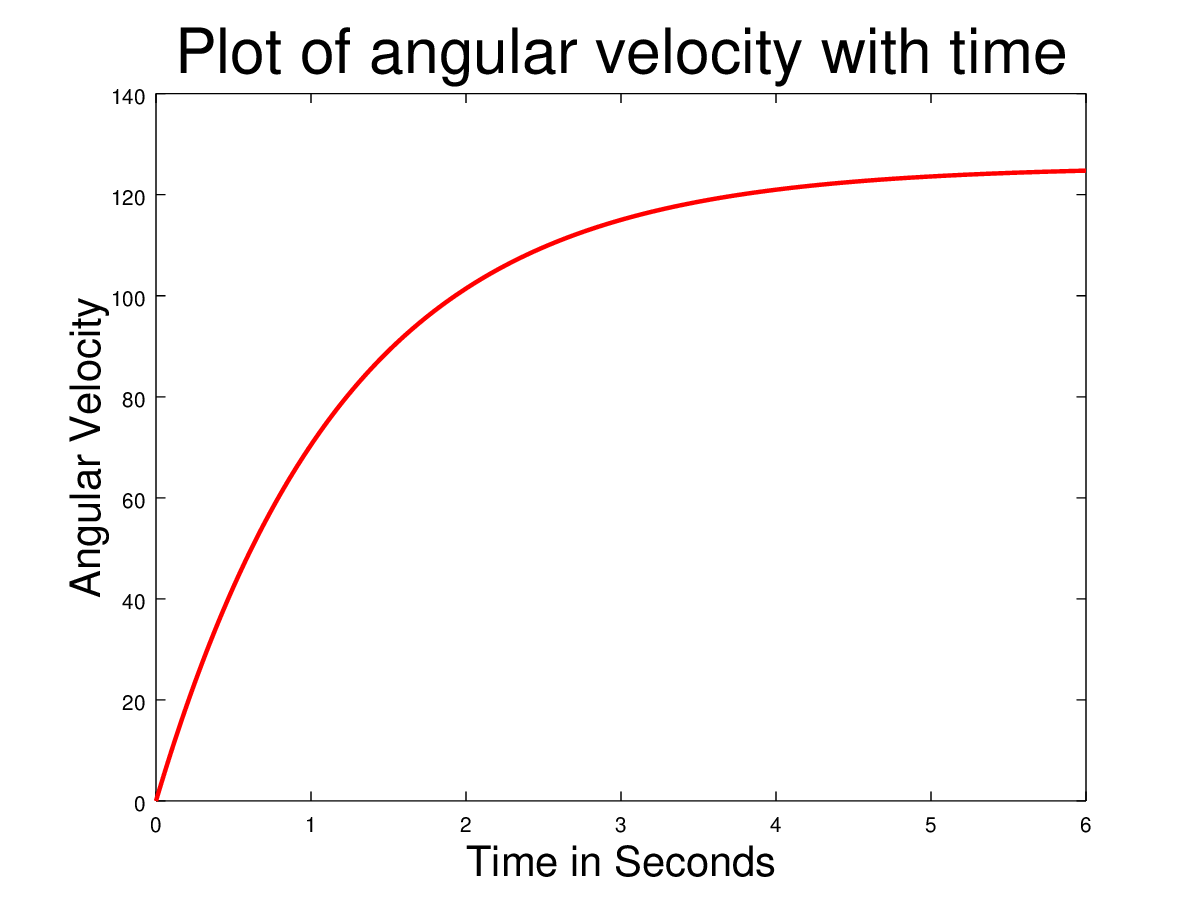}}
  \caption{Simulation results for a 3-phase permanent magnet motor. The torque function for one phase is shown in the left-top figure. The other torque functions are all circularly rotated functions (rotated by multiples of $2\pi/3$). The top-right plot shows all the currents in the three coils, while the bottom-left figure shows a zoomed in view of the currents. The angular velocity is shown in the bottom-right figure.\label{fig:normal}}
\end{sidewaysfigure}

\begin{sidewaysfigure}
  \centering
  \subfloat{\includegraphics[scale=0.45]{nonlinearity.png}} \quad
  \subfloat{\includegraphics[scale=0.45]{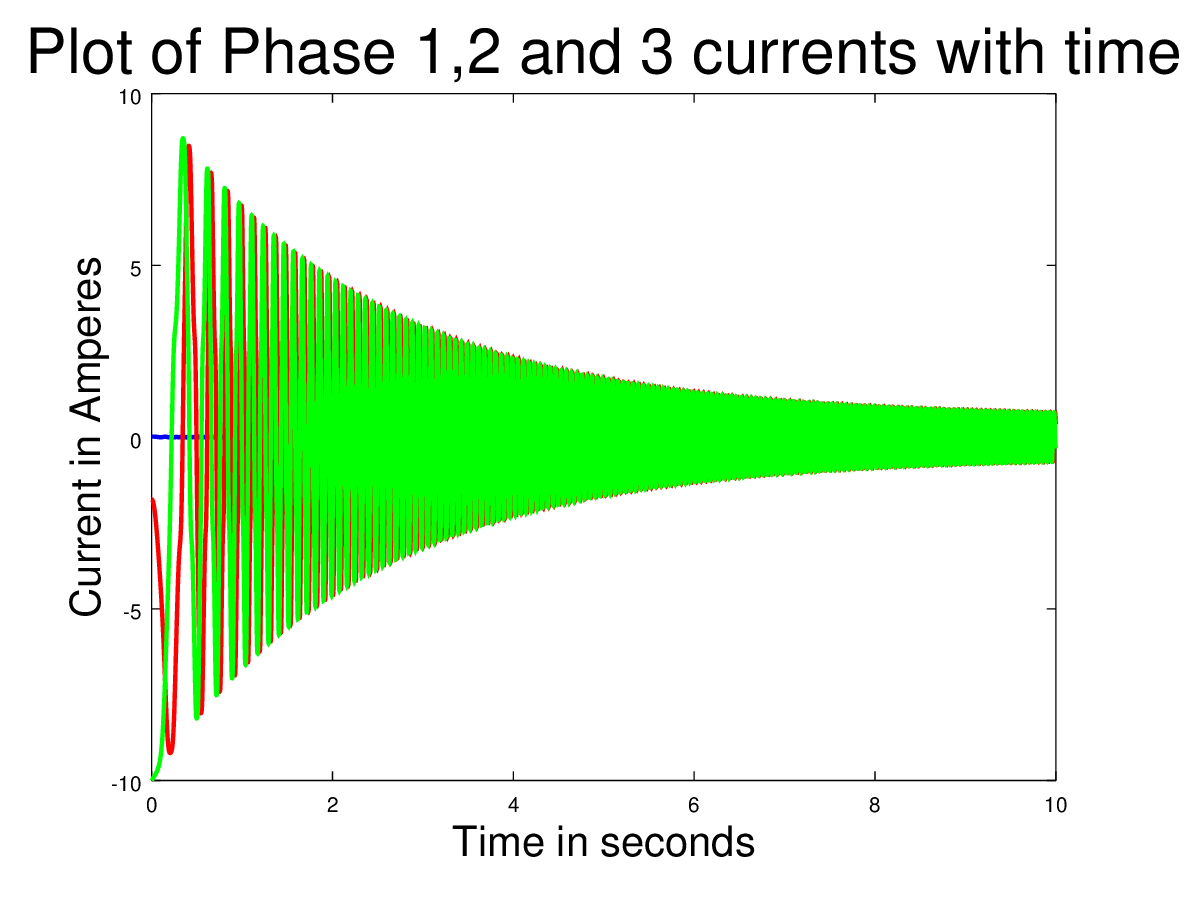}}\quad
  \subfloat{\includegraphics[scale=0.45]{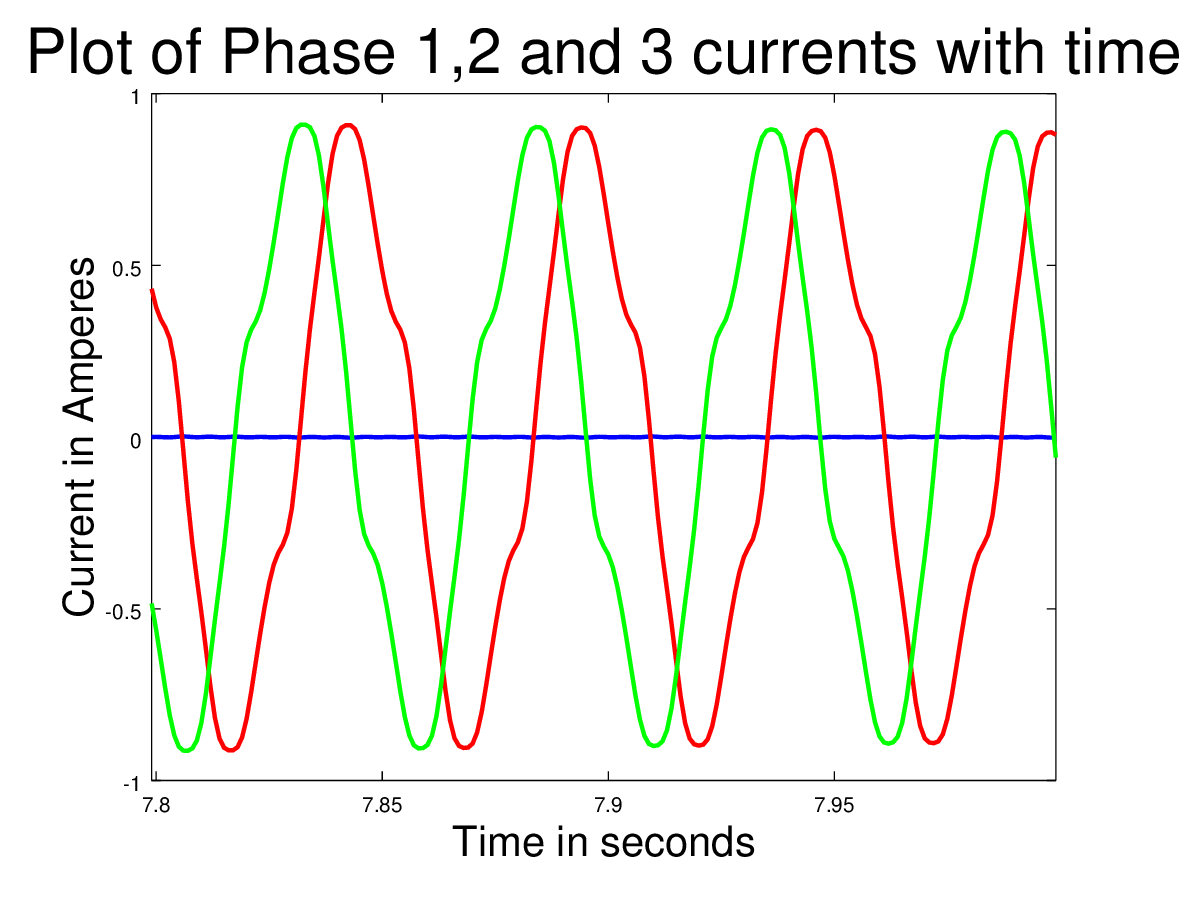}}\quad
  \subfloat{\includegraphics[scale=0.45]{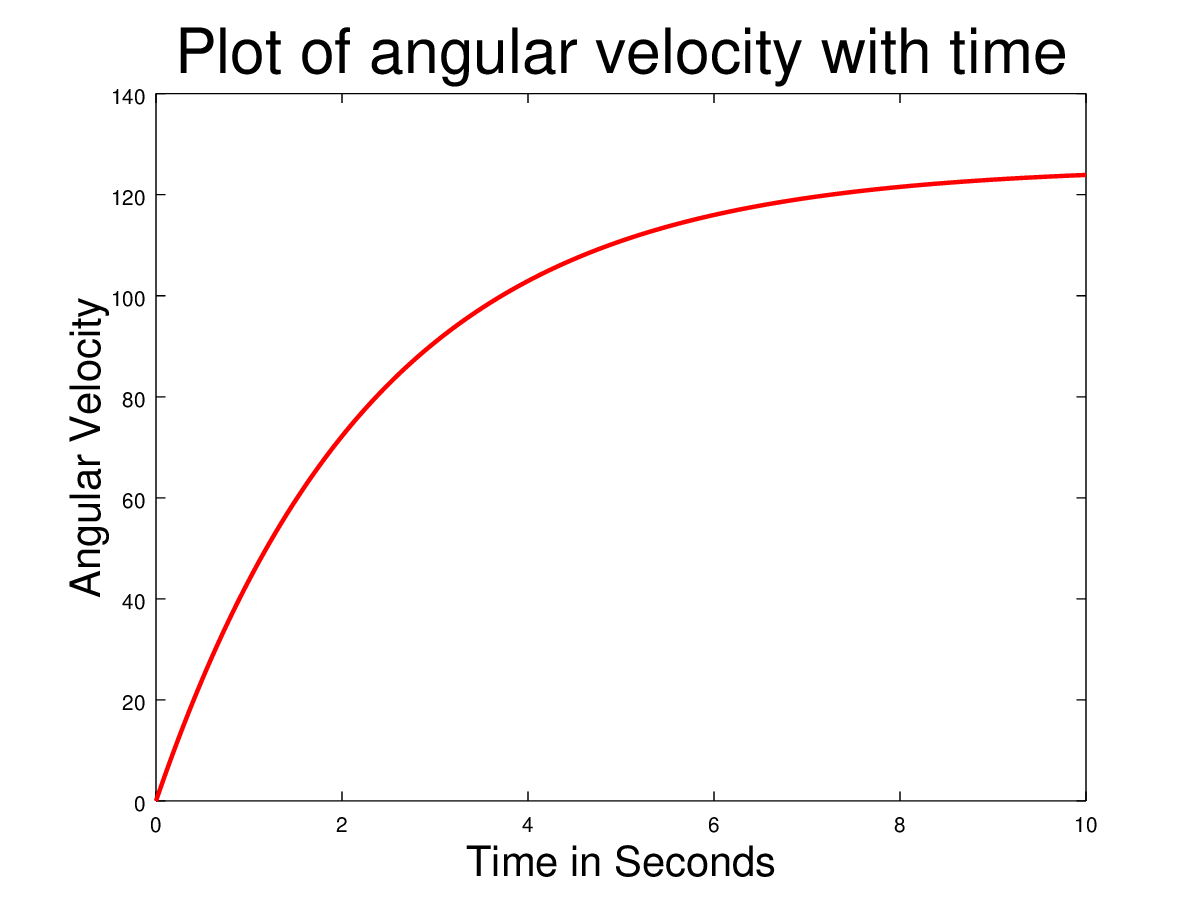}}
  \caption{Simulation results for a 3-phase permanent magnet motor, with one non-conducting faulty coil. The torque function for one phase is shown in the left-top figure. The other torque functions are all circularly rotated functions (rotated by multiples of $2\pi/3$). The top-right plot shows all the currents in the three coils, while the bottom-left figure shows a zoomed in view of the currents. The angular velocity is shown in the bottom-right figure.\label{fig:onefault}}
\end{sidewaysfigure}

\begin{sidewaysfigure}
  \centering
  \subfloat{\includegraphics[scale=0.45]{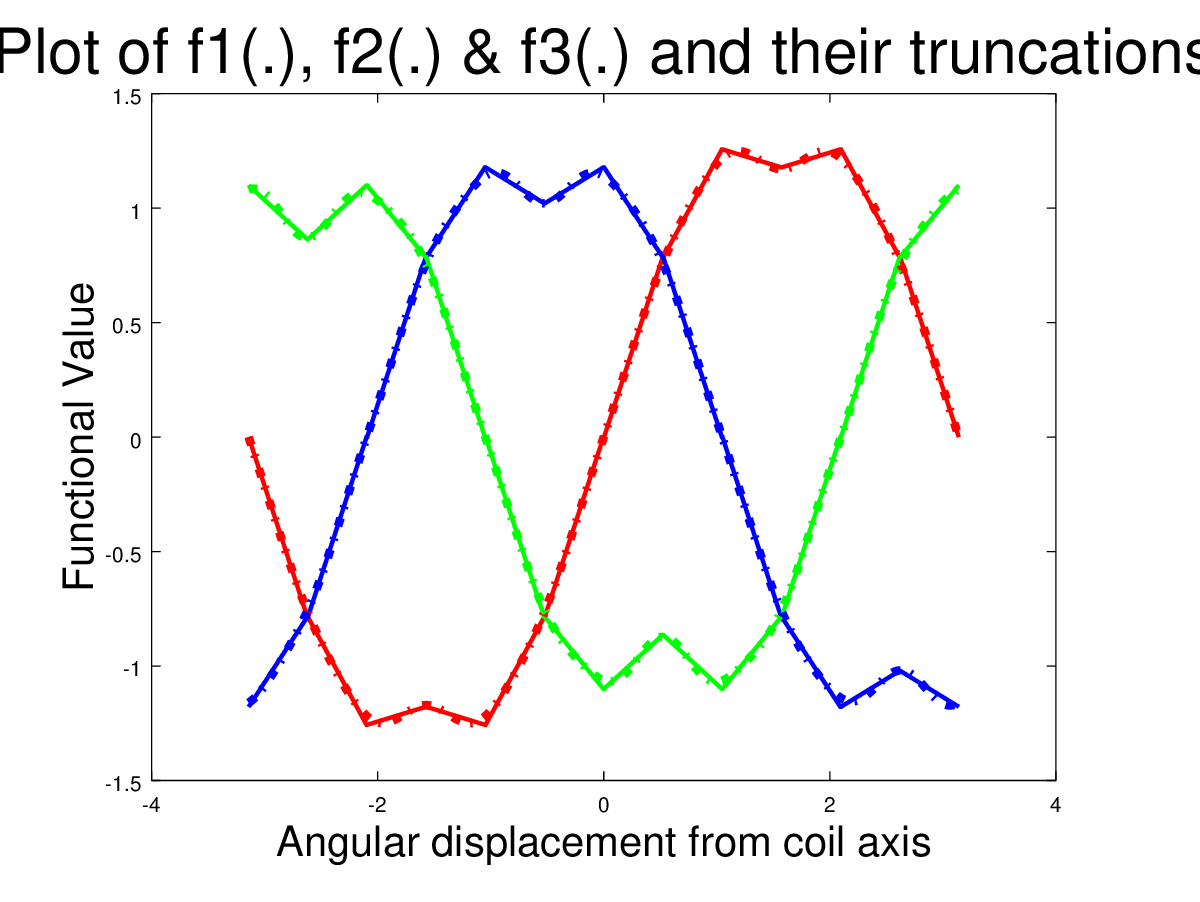}} \quad
  \subfloat{\includegraphics[scale=0.45]{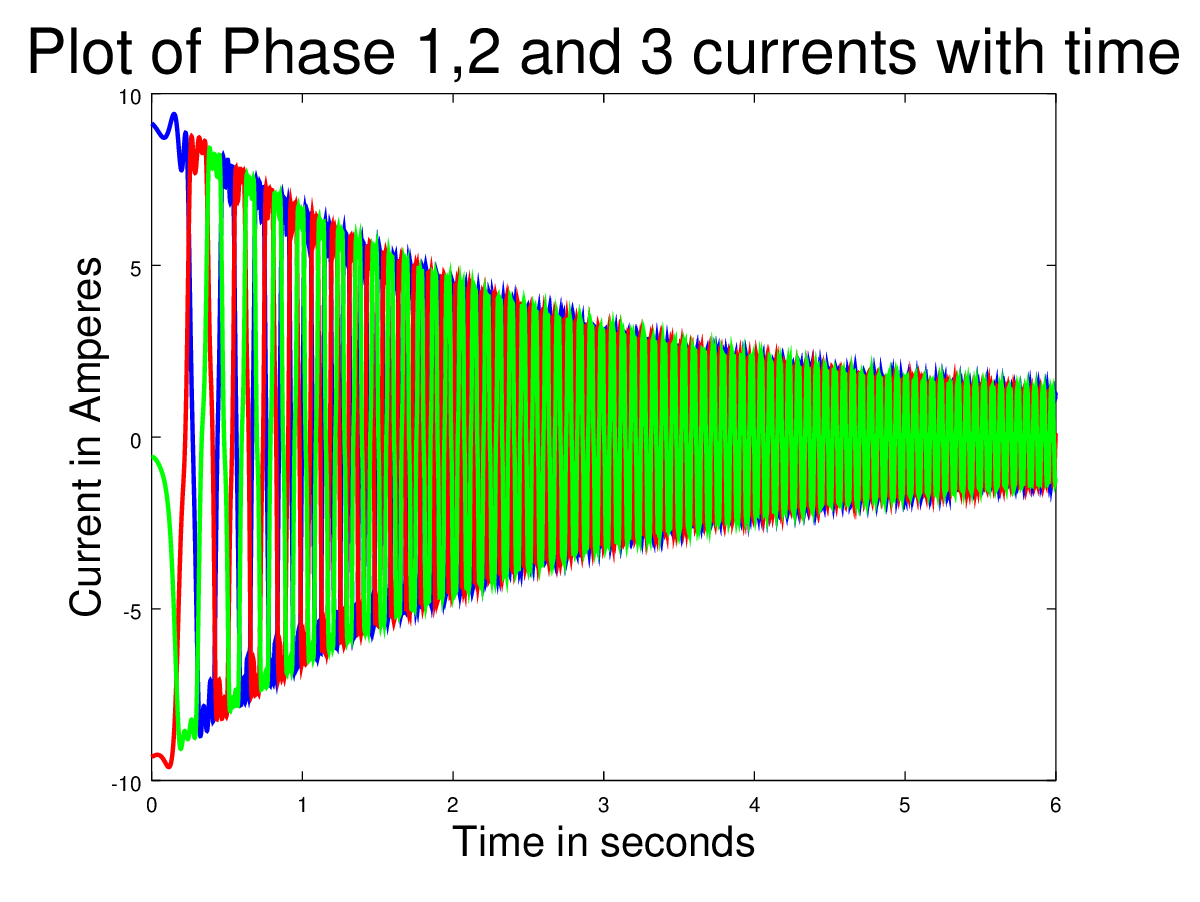}}\quad
  \subfloat{\includegraphics[scale=0.45]{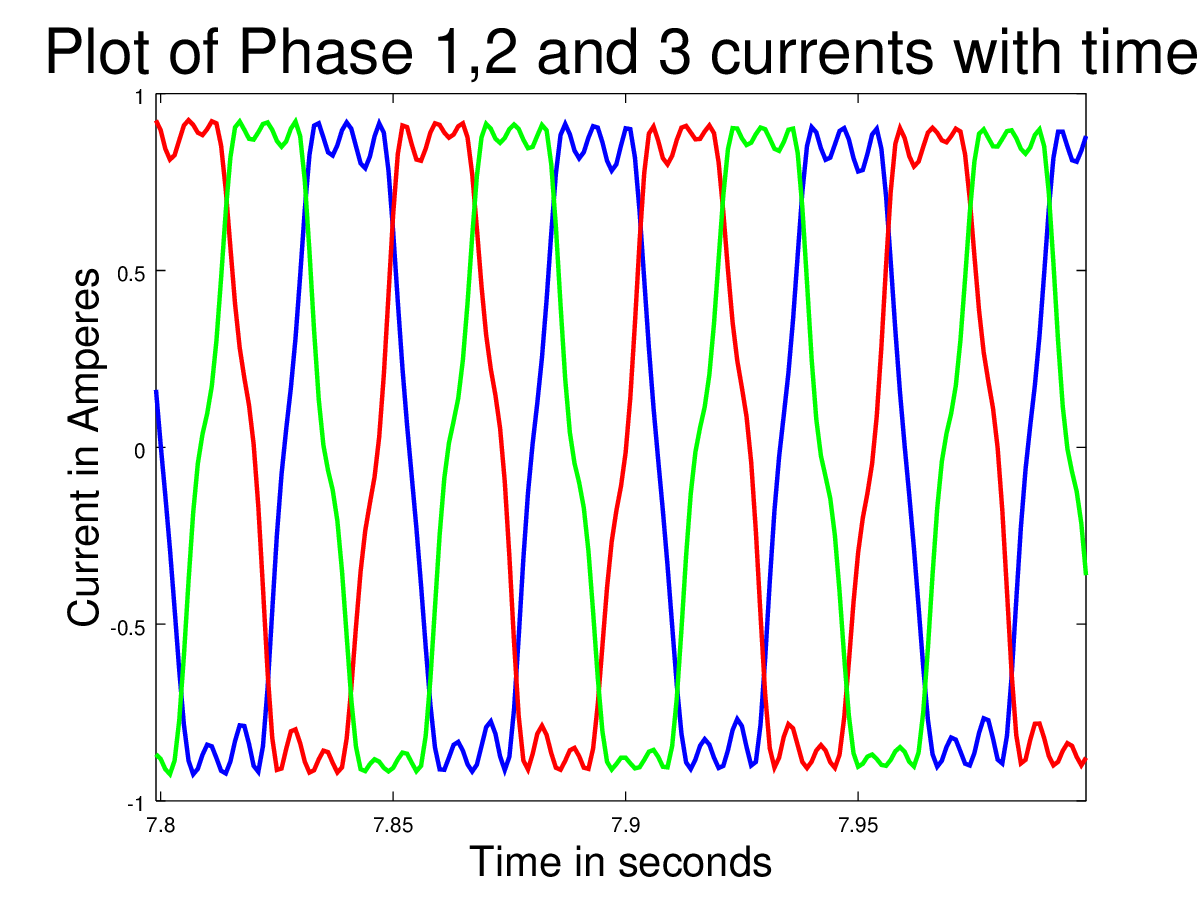}}\quad
  \subfloat{\includegraphics[scale=0.45]{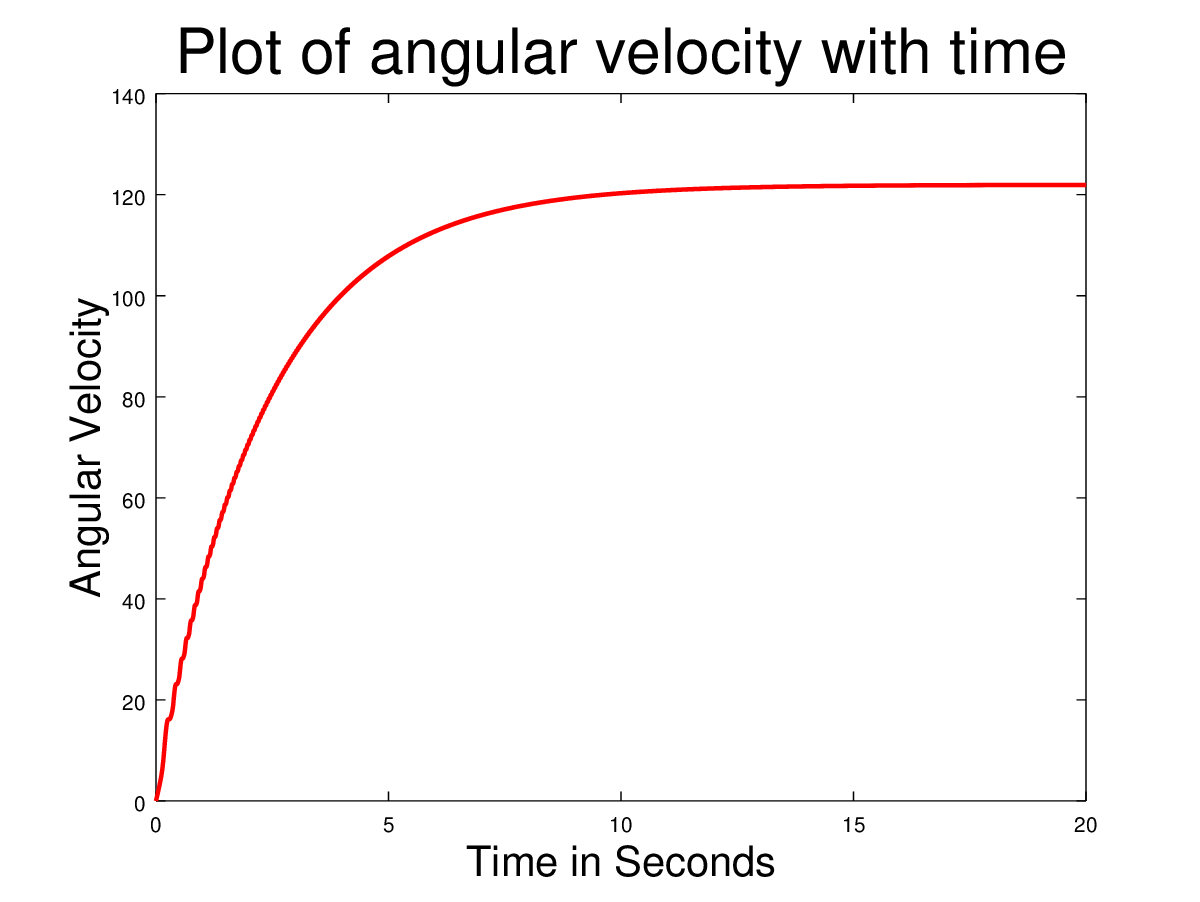}}
  \caption{Simulation results for a 3-phase permanent magnet motor with unbalanced torque functions. The torque functions for all three  phases are shown in the left-top figure. The top-right plot shows all the currents in the three coils, while the bottom-left figure shows a zoomed in view of the currents. The angular velocity is shown in the bottom-right figure.\label{fig:unbalanced}}
\end{sidewaysfigure}

The simulations carried out here serve as a proof of concept for the presented method. The torque function, which is nonlinear, the motor parameters and the torque inputs are all representative; the actual parameters for a practical motor need to be identified. For the purpose of clarity, it is assumed that the back-emf is very low in these cases and hence have been ignored. Moreover, it is also assumed that the switching voltage waveform is appropriately generated (as per the discussion in a previous section) to induce the required currents. The optimization was done using CVXPY \cite{diamond2016cvxpy}.\\\\
The first set of figures in \ref{fig:normal} depict the torque function, the current generated as a result of applying appropriate control voltages, the speed curve with respect to time and a zoomed-in view of the current waveform once the speed has reached sufficiently close to the desired speed. Moreover, it is assumed that torque functions for the three coils of the motor are symmetric barring the constant angular displacement. The current is limited to 10 Amperes, and hence it can be seen that the currents never exceed this limit. The opposing torque input has a constant value of 3 Nm. Moreover, the current waveform is not sinusoidal, which is the reason the motor speed increases as per a linear ODE. The second set of figures in \ref{fig:onefault} is the case where a coil is faulty and does not conduct any current. It can be seen that the currents in the other coils increase to compensate for the input torque. Moreover, the response time in this case is almost 10 seconds, as compared to the normal case where it was 6 seconds. Note that the torque function and all other parameters are the same here, as in the first case. In the third case \ref{fig:unbalanced}, the torque function for the three coils are not symmetric, indicating possible faults in design and manufacturing the motor. The non-linearities are shown in the same set of figures. It can be seen that the proposed algorithm handles this case with a response time of almost 6 seconds. Note that the current waveforms in different coils are also not symmetric, unlike the first case. The fourth set of figures represent the results for a balanced five phase motor. It can be seen that for the same nonlinearity and the same current bounds, the response time is almost 3 seconds. This is natural since 5 coils can produce a lot more torque. Also note that the waveforms of the currents in the five coils are quite different as compared to the first case.  

\begin{sidewaysfigure}
  \centering
  \subfloat{\includegraphics[scale=0.45]{nonlinearity.png}} \quad
  \subfloat{\includegraphics[scale=0.45]{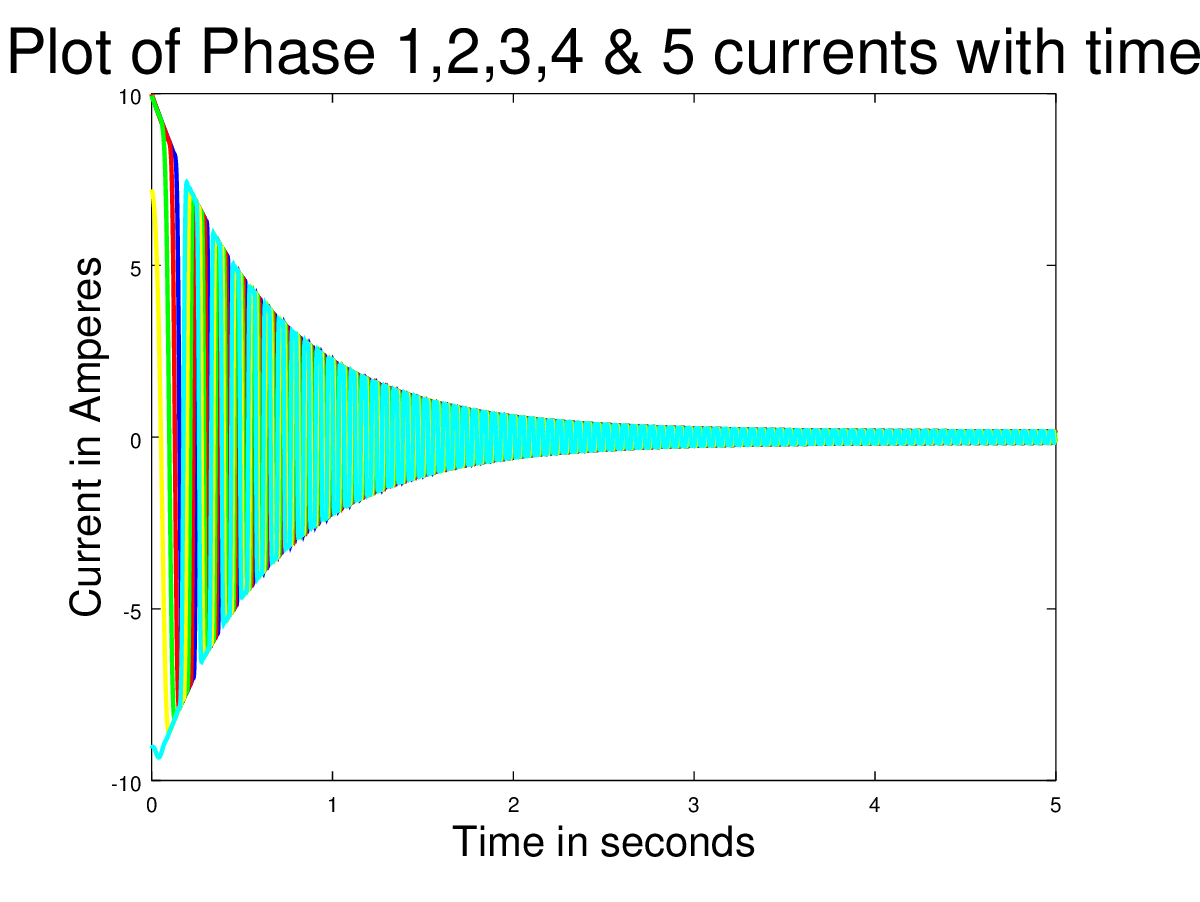}}\quad
  \subfloat{\includegraphics[scale=0.45]{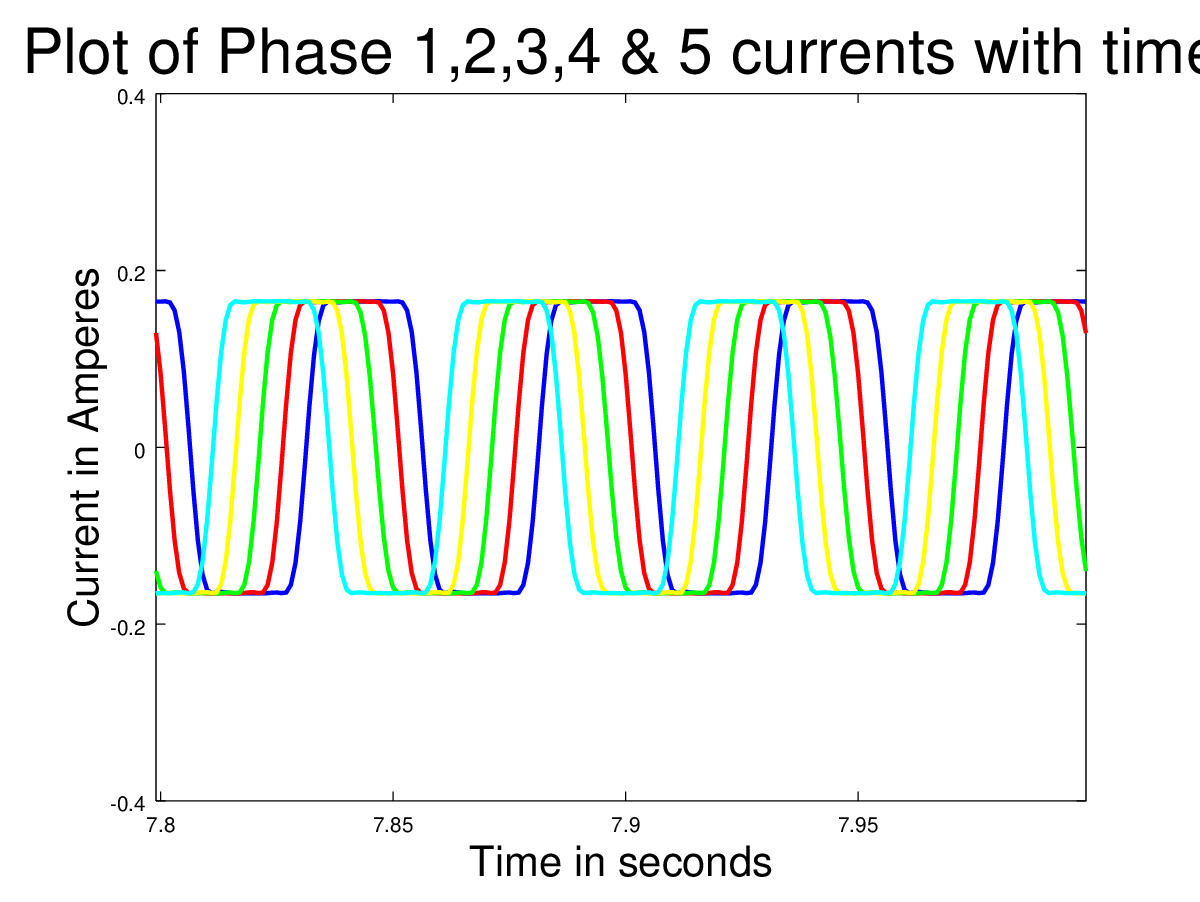}}\quad
  \subfloat{\includegraphics[scale=0.45]{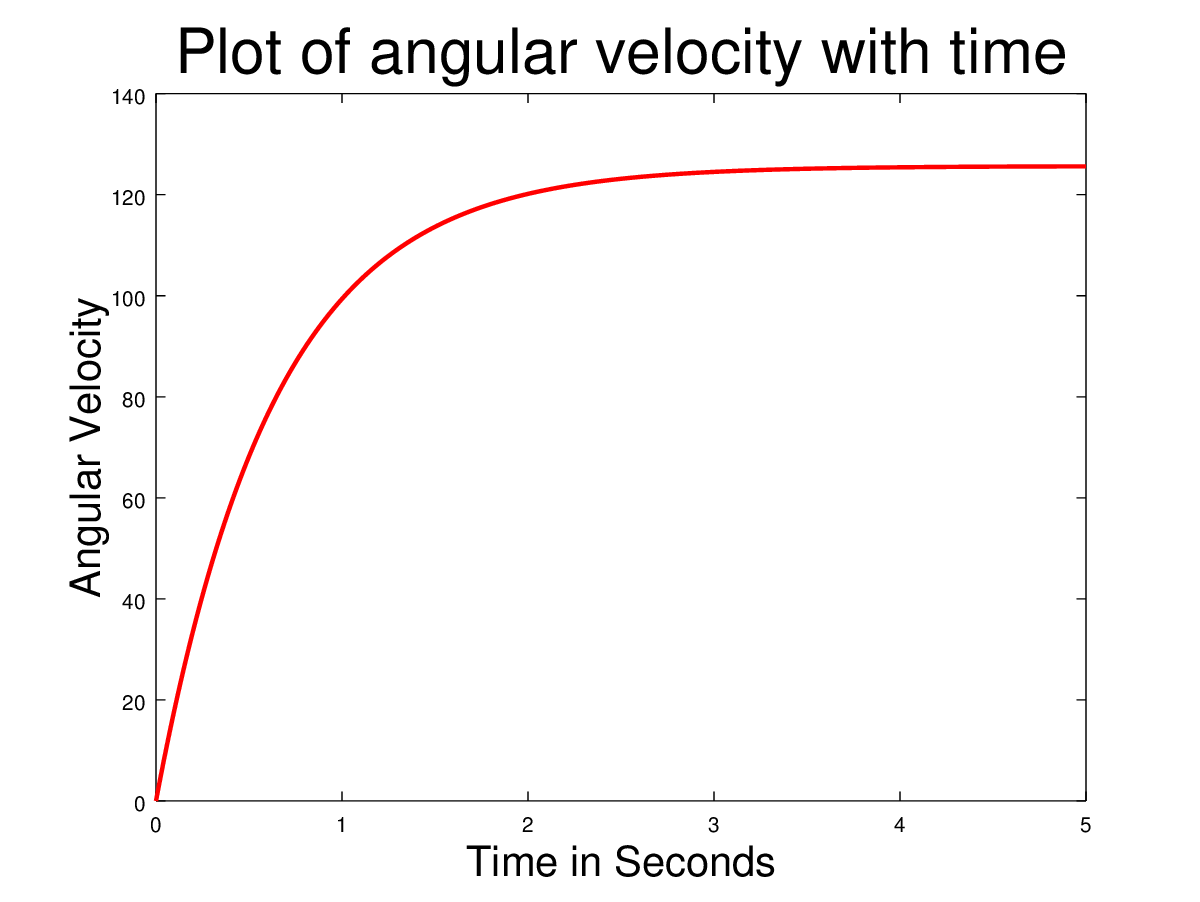}}
  \caption{Simulation results for a 5-phase permanent magnet motor, where the coils are not equally spaced out. The torque function for one phase is shown in the left-top figure. The other torque functions are all circularly rotated functions (rotated by multiples of $\pi/5$, instead of $2\pi/5$). The top-right plot shows all the currents in the five coils, while the bottom-left figure shows a zoomed in view of the currents. The angular velocity is shown in the bottom-right figure. \label{fig:fivephase}}
\end{sidewaysfigure}

\section{Conclusions}
In this paper, a method of determining a nonlinear control algorithm for a permanent magnet motor with current and voltage constraints was devised. The dynamical system model for the motor was standard except for the general functional dependence on rotor angle of the torque generated by a current in one of the stator coils. The key idea was based on a sinusoidal representation of the closed loop control, usage of positivity condition of trigonometric polynomials and convex semidefinite programming. This way the system was reduced to a linear first order system, which was then easy to analyze. The system was also analyzed under noisy feedback and uncertainties in model parameters, and certain explicit performance guarantees were given. Simulations corroborated the idea proposed in this paper.

\bibliographystyle{alpha}
\bibliography{resfs}
\end{document}